\documentclass[prd]{revtex4}
\usepackage{color,graphicx}
\usepackage{epsf}

\usepackage{amsbsy}
\usepackage{amssymb}
\usepackage{amsmath}
\begin{document}

\title{r-modes in low temperature colour-flavour-locked superconducting quark stars}

\author{N. Andersson$^1$, B. Haskell$^1$ and G.L. Comer$^2$ }
\address{$^1$School of Mathematics, University of Southampton, Southampton, UK\\
$^2$Department of Physics and Center for Fluids at All Scales, Saint Louis University, St Louis, USA }

\date{\today}

\voffset 0.5 truein

\def\be{\begin{equation}}
\def\ee{\end{equation}}
\def\bea{\begin{eqnarray}}
\def\eea{\end{eqnarray}}
\def\bear{\begin{eqnarray}}
\def\eear{\end{eqnarray}}
\def\beq{\begin{eqnarray}}
\def\eeq{\end{eqnarray}}
\def\n{{\rm n}}
\def\x{{\rm x}}
\def\s{{\rm s}}
\def\N{{\rm N}}
\def\S{{\rm S}}
\def\mun{{\mu_\n}}
\def\mus{{\mu_\s}}

\begin{abstract}
We present the first  multi-fluid analysis of a dense neutron star core with a deconfined colour-flavour-locked superconducting quark component.
Accounting only for the condensate and (finite temperature) phonons, 
we  make progress by taking over results for superfluid $^4$He. The resultant two-fluid model accounts for a number of additional viscosity coefficients 
(compared to the Navier-Stokes equations) and we show how they enter the dissipation analysis for an oscillating star. 
We provide simple estimates for the gravitational-wave driven r-mode instability, demonstrating that 
the various phonon processes that we consider are not effective damping agents. Even though the results are likely of little direct astrophysical 
importance (since we  consider an overly simplistic stellar model)  our analysis represents significant technical progress,
laying the foundation for more detailed numerical studies and  preparing the ground 
for the inclusion of additional aspects (in particular associated with kaons) of the problem.
\end{abstract}

\maketitle

\section{Introduction}

The state of matter at extreme densities continues to be an issue of vigorous investigation. The problem is
complicated, not only from the theoretical point of view, but also by the fact that
laboratory experiments are restricted. For example, while colliders like RHIC at Brookhaven, GSI in Darmstadt and the LHC at CERN probe
 hot quark-gluon plasmas they will never be able to explore the cold, extreme high density, region of the QCD phase diagram.
In order to test our understanding of the relevant physics we need to turn to astrophysics, and the
dynamics of compact stars. In fact, ``neutron stars'' represent unique laboratories of such extreme
physics. With core densities  reaching about one order of magnitude beyond nuclear saturation,
they are likely to contain exotic states of matter like hyperon phases with net strangeness and/or
deconfined quarks. It is well-established that these states of matter should exhibit
superfluidity/superconductivity at the relevant temperatures (neutron stars are born with temperatures $\sim 10^{12}$~K
and rapidly cool below $\sim 10^9$~K). Moreover, observed radio pulsar glitches provide strong evidence for the
presence of a partially decoupled superfluid component in these systems. The modelling of the dynamics of
these, potentially \underline{very} complex, objects presents a serious challenge.

A key aspect of the problem concerns the fact that a superfluid system has additional
dynamical degrees of freedom. This is well-known from experiments on laboratory systems
like $^4$He, which exhibit a second sound associated with thermal waves \cite{khalatnikov,putterman}. Analogous ``superfluid''
modes have been studied in detail for superfluid neutron-proton-electron mixtures relevant for the outer core
of a neutron star \cite{epstein,mendell,ac01}. For simplicity, these studies have almost exclusively ignored thermal effects (the work in \cite{gusakov} is a notable exception). 
While this is a useful first approximation it is clear that
the zero temperature assumption must be relaxed in a realistic model. Basically, due to the density
dependence of the various superfluid pairing gaps (see \cite{NPA} for a guide to the literature), there will \underline{always} be regions
in a neutron star where thermal effects are important (in the vicinity of the critical density at which the
phase transition occurs). Understanding the nature of these transition regions, and their effect on various 
aspects of neutron star dynamics, is one of the main challenges for  research in this area. 

In this paper, we describe a first attempt at modelling thermal dynamics in a superfluid neutron star.
We focus on stars with a colour-flavour-locked (CFL) superconducting quark core \cite{CFL}
at finite temperatures. This is an interesting problem for several reasons. First of all, the simplest
possible model for this system considers a quark condensate coupled to a gas of phonons.
This problem is analogous to $^4$He at low temperatures, and hence we can bring our recent
dissipative two-fluid model \cite{helium} to bear on it (more or less directly). The lessons we learn
from this exercise should inform the development of finite temperature models for the superfluids
in the outer core and the neutron star crust.
Secondly, even though there have been discussions of the dynamics of the different
CFL phases, in particular in the context of the gravitational-wave driven r-mode instability (see \cite{jaikumar} for references),
the superfluid aspects have (as far as we are aware) not previously been accounted for.
Our discussion begins to address the relevance of the additional degrees of freedom in these systems, and  provides some
insight into the nature of the different ``fluids'' involved.

As an application with immediate astrophysical relevance, we will work out the inertial r-modes
and estimate the relevant viscous damping rates for the
``simplest'' model of CFL matter. We consider the ``cool'' regime where the
temperature is significantly below all the quasiparticle energy gaps. In this regime,
dissipation may mainly occur due to phonon interactions.
One reason for considering this model is that there are results in the literature
for both bulk- and shear viscosity \cite{manuel1,manuel2} as well as the mutual friction associated with superfluid vortices \cite{mana1}.
Of course,
the simple ``condensate plus phonon'' model that we  consider is not the whole story. It should apply at asymptotically high densities,
but may not be the true ground state at lower densities.
The discussion in \cite{alford1,alford2,alford3} adds extra dimensions to the problem
by considering the bulk viscosity due to kaons (allowing for
flavour-changing processes), which will be present at
higher temperatures. At first sight, this mechanism will only be relevant for
very hot stars ($T>1$~MeV~$\sim 10^{10}$~K) since it assumes that there is a thermal population of kaons.
Below the critical energy where the kaons appear
the contribution to the bulk viscosity is exponentially suppressed and may not be that important. However,
in more recent work on the so-called CFL-K$^0$ phase \cite{alford2}, it is argued that the main low temperature
mechanism involves condensed kaons. This is important since the kaon condensate will remain present as $T \to 0$. The upshot of this is
that
the problem requires a ``multi-fluid'' analysis
at all temperatures. The simplest model would  have three components;  the quark condensate, the kaon condensate and
finite temperature excitations (phonons and thermal kaons). This problem is  more involved than the case that we focus on here.
Nevertheless, our results provide an essential starting point for investigations into the dynamical role of the kaons.
Most importantly, by providing the ``hydrodynamics'' view of the problem we illustrate the 
input needed to study various dynamical scenarios. This should stimulate further discussion between 
experts on different aspects of this multi-faceted problem, as required to make progress in the future.

\section{Flux-conservative two-fluid model}

We  consider a deconfined quark system  system that contains a
single CFL superconducting condensate and a phonon gas. Formally, this problem is  identical to
that for $^4$He at low temperatures. Hence, we can take as our starting point
the recent discussion of superfluid Helium \cite{helium}, which is based on the
convective variational approach to multi-fluid dynamics \cite{prix,monster,livrev} and
which we know is in one-to-one correspondence with the orthodox formulation
developed by, in particular, Khalatnikov \cite{khalatnikov}.

The variational model takes as its starting point an energy functional $E$, representing the equation of state. 
This energy determines the relation between the various fluxes and
the associated canonical momenta. In the present case, 
we distinguish between the massive particles in the system, with number density $n$ and flowing with
$ n v^\n_i$ from a massless entropy component with number density $s$ and
flux $sv^\s_i$. The latter represents the phonons (which are treated in the fluid approximation, i.e., we assume that 
there exists a suitable defined ``average'' transport velocity for each of the two components in the system) The two
momentum densities then follow from~\footnote{Throughout this paper we use a coordinate basis to represent tensorial relations.
That is, we distinguish between co- and contra-variant objects, $v_i$ and $v^i$, respectively.
Indices, which range from 1 to 3, can be raised and lowered with the (flat space) metric $g_{ij}$, i.e.,
$v_i = g_{ij} v^j$. Derivatives are expressed in terms of the covariant derivative
$\nabla_i$ which is consistent with the metric in the sense that $\nabla_i g_{kl} = 0$.
This formulation has great advantage when we want to discuss the geometric nature of the
different dissipation coefficients.}
\be
   \pi_i^\n = mn v_i^\n - 2 \alpha w_i^{\n\s} \ ,
\ee
and
\be
   \pi^\s_i = 2 \alpha w_i^{\n\s} \ ,
\ee
where $w_i^{\n\s} = v^\n_i-v^\s_i$ and $\alpha$ represents the entropy
entrainment \cite{heat}. It has been assumed that $E=E(n,s,w^2)$, as expected in an isotropic system, and we have defined
\be
\alpha = \left. {\partial E \over \partial w^2} \right|_{n,s} \ ,
\ee
omitting the indices on $w$ for clarity.

As discussed in \cite{helium,monster}, the associated momentum equations can be
written
\be
   f_i^\n = \partial_t \pi_i^\n + \nabla_j (v_\n^j \pi_i^\n +
   D^{\n j}_{\ \ i} ) + n \nabla_i \left( \mu_\n - \frac{1}{2} m v_\n^2
   \right) + \pi_j^\n \nabla_i v_\n^j  \ , \label{eulern}
\ee
and
\be
f_i^\s = \partial_t \pi_i^\s + \nabla_j (v_\s^j \pi_i^\s + D^{\s j}_{\ \ i} )
+ s \nabla_i T + \pi_j^\s \nabla_i v_\s^j \ , \label{eulers}
\ee
where $\mu_\n$ is the matter (quark) chemical potential;
\be
\mu_\n = \left( { \partial E \over \partial n} \right)_{s, w^2} \ ,
\ee
and we have used the fact that the temperature corresponds to the entropy chemical potential \cite{heat};
\be
T= \mu_\s = \left( { \partial E \over \partial s} \right)_{n, w^2} \ .
\ee
In these expressions,
$D^\x_{ij}$ represent the viscous stresses while the ``forces'' $f_i^\x$ allow for momentum
transfer between the two components. In the following we will assume that the
system is isolated, which means that $f_i^\n+f_i^\s=0$. As we will discuss later, the force terms can also be used to account for ``external'' forces like gravity.

In the present context, when we are dealing with a single matter quantity, we will have \cite{monster} 
\be
   \partial_t n + \nabla_j (n v_\n^j ) = \Gamma_\n = 0 \ ,
\ee
as there is no particle creation/destruction.
At the same time the entropy can increase, so we have
\be
   \partial_t s + \nabla_j (s v_\s^j ) = \Gamma_\s \ge 0 \ .
\ee
We also have \cite{monster}
\be
   T \Gamma_\s = - f_i^\n w_{\n\s}^i - D^j_{\ i} \nabla_j v_\s^i -
   D^{\n j}_{\ i} \nabla_j w_{\n\s}^i \ , \label{TGs}
\ee
where
\be
D_{ij} = D_{ij}^\n + D_{ij}^\s \ .
\ee

For an isolated system we  know that, if we impose the
superfluid constraint of irrotationality the number of
dissipation coefficients reduces significantly. As discussed in \cite{helium},
we have
\be
- D_{ij} =  g_{ij} (\hat{\zeta}^{\n} \nabla_l j^l + \zeta \Theta_\s)
+ 2 \eta \Theta^\s_{ij} \ . \label{Dij}
\ee
where we have defined $j^i = n w^i_{\n \s}$.
We also have;
\be
 \frac{1}{n} \left( f_i^\n - \nabla_l D^{\n l}_{\ \ i} \right) = \nabla_i \Psi
\label{phi1}
\ee
with
\be
\Psi =  \hat{\zeta}^{\n\n} \nabla_l j^l + \hat{\zeta}^{\n} \Theta_\s
\label{phinal} \ .
\ee
At this point, only four dissipation coefficients remain in the problem.

In the above expressions, we have used the standard decomposition;
\be
\nabla_i v^\s_j = \Theta^\s_{ij} + \frac{1}{3} g_{ij} \Theta_\s +
\epsilon_{ijk} W_\s^k
\ee
in terms of, the expansion
\be
\Theta_\s = \nabla_j v_\s^j \ ,
\label{exp}\ee
the trace-free shear
\be
\Theta^\s_{ij} = \frac{1}{2} \left( \nabla_i v^\s_j + \nabla_j v^\s_i -
\frac{2}{3} g_{ij} \Theta_\s \right) \ ,
\label{sh}\ee
and the ``vorticity''
\be
W_\s^i = \frac{1}{4} \epsilon^{ijk}( \nabla_j v^\s_k - \nabla_k v^\s_j) \ .
\ee
We  use analogous expressions for gradients of the relative velocity.
The definition
of the various quantities should be obvious from the constituent indices.

However, since the system that we consider is not irrotational we need to 
consider relaxing the assumptions on the 
dissipation coefficients. This involves making some subtle decisions.  In the case of 
an irrotational flow it is natural to assume that  $f_\n^i=0$. 
When the superfluid rotates, and vortices are present, the force will not vanish.
It is necessary to account for  dissipation due to, for example, the  scattering of phonons off of the vortex cores.
The standard approach to the rotating problem is to add in this ``mutual friction'' force, 
keeping the other dissipative terms as in the irrotational case. This strategy ignores a number of 
dissipative terms that would, at least in principle, be allowed in the equations of motion \cite{helium,monster}. 
The role of these additional terms has not yet, as far as we are aware,  been investigated. 

Ignoring the potential relevance of most of the additional dissipation channels in the irrotational case, we
simply account for the presence of vortices by
i) assuming that \eqref{eulern} represents fluid elements with both a condensate and a
smooth-averaged vorticity arising from the vortices, and ii) accounting for the vortex mediated
mutual friction by allowing a force \cite{trev}
\be
f_i^\mathrm{mf} = \mathcal{B}' \rho_\n n_v \epsilon_{ijk} \kappa^j w_{\n\s}^k
+ \mathcal{B} \rho_\n n_v \epsilon_{ijk} \epsilon^{klm} \hat{\kappa}^j
\kappa_l w_m^{\n\s} \ , \label{mf}
\ee
to act on the particles (with a balancing force affecting the excitations).
Here $n_v$ is the vortex area density and the vector $\kappa^i = \kappa \hat{\kappa}^i$ (the hat represents a unit vector)
is aligned with the rotation axis and has magnitude $\kappa= h/2m$. Since we will only consider
 the effect of phonon scattering off of vortices, we expect to be in the weak mutual friction regime
where $\mathcal{B}' \ll \mathcal{B}$. This means that the first term in the force \eqref{mf} can be ignored, 
leaving only the second, dissipative, contribution.

\section{Rotating equilibrium models}

In order to set the stage for the discussion of r-modes, we need to provide a suitable rotating equilibrium configuration.
To do this, we  note that the two components flow together (there is not heat flux) when the system is in thermal equilibrium.
This means that $w_i^{\n\s}=0$, which implies that $\pi_i^\s=0 $. Thus, it follows from the entropy momentum equation \eqref{eulers}
that we must have
\be
s \nabla_i T = 0 \quad \longrightarrow \quad T = \mathrm{constant} \ .
\ee
This is trivial; if the system is isothermal then there will be no heat flux.

Rewriting the remaining momentum equation \eqref{eulern}, making use of the continuity equation, we find (after accounting for the gravitational force
as the gradient of the gravitational potential $\Phi$)
\be
\partial_t v_i^\n + v_\n^j \nabla_j v^\n_i + \nabla_i (\tilde{\mu}_\n + \Phi) = 0 \ ,
\ee
where $\tilde{\mu}_\n = \mu_\n/m$. Using the Lie-derivative along the flow, $\mathcal{L}_{v_n}$, we can rewrite this equation as
\be
\partial_t v^\n_i + \mathcal{L}_{v_n} v_i^\n +  \nabla_i \left(\tilde{\mu}_\n + \Phi - { 1 \over 2} v_\n^2 \right) = 0 \ .
\ee
Restricting ourselves to stationary models we have 
\be
\partial_t v_i^\n = 0 \ .
\ee
Next,  uniform rotation implies
\be
v_\n^i = \Omega_\n \mathbf{e}_\varphi^i \ ,
\ee
and, for axisymmetric models, it then follows that
\be
\mathcal{L}_\varphi v_i^\n = 0 \ .
\ee
Hence, the required equilibrium models are determined from the usual Bernoulli-type equation
\be
\tilde{\mu}_\n + \Phi - { 1 \over 2} v_\n^2 = \mathrm{constant} \ .
\ee
This analysis shows that a rotating configuration can be obtained in the usual way.
Once we assume thermal equilibrium, we are dealing with a single-fluid problem.

\section{Linear perturbations}

Our main aim is to develop the tools required to make quantitative estimates for the r-mode 
instability in neutron stars with a CFL core. In doing this, we want to 
account for the fact that CFL matter requires a multi-fluid description. The analysis 
proceeds in three steps. First we need to formulate the linear perturbation problem 
for the system. This step is interesting because, as far as we are aware, this is the first time 
that perturbations of a matter plus massless entropy system have been considered in an 
astrophysical context. The results provide useful insights into finite temperature 
superfluid dynamics, and should be relevant in a broader context. 
The second step corresponds to determining the 
pulsation modes of the system, in this case the r-modes, and finally we need to estimate the damping timescales 
associated with the different dissipation channels. In the discussion below, we only work out the 
first of these steps in  detail. In order to obtain useful estimates for the r-mode instability
we simplify last two steps by considering a uniform density model. This has the advantage that 
the r-mode solution is simple, and we can evaluate the dissipation rates analytically. 
In a sense, we do not expect this approximation to be too bad because it is well known that 
the density profile for a canonical $1.4M_\odot$/10~km strange star (described by the MIT bag model)
is almost flat (see \cite{alcock}).  

Of course, we really need to develop a consistent
model for a realistic equation of state. This is essential if we want to  study 
hybrid stars, where the CFL phase is only present in the  core. However, the 
problem is complicated by the fact that the bulk viscosity damping of the r-modes 
requires the perturbations to be worked out to second order in the slow-rotation approximation \cite{review}. 
This necessitates a numerical solution, which makes
some of the qualitative behaviour of the results less clear. Another complicating factor is the well-known fact that 
it only makes sense to use a realistic equation of state in a fully general relativistic analysis (see \cite{hyperon} for
discussion). The r-modes in single component relativistic stars have been studied \cite{kl1,kl2,jr1,jr2}, but there has not yet been any 
serious analysis of the corresponding multi-fluid problem (although see \cite{greg}). A focussed effort in this direction 
should be encouraged.

\subsection{The non-dissipative problem}

In order to work out the r-mode solutions, it is advantageous to work in a rotating frame.
Focussing, for the moment, on the non-dissipative equations we find that, in a frame
 rotating uniformly with $\Omega^i$ we have
\be
   \partial_t \pi_i^\n + \nabla_j (v_\n^j \pi_i^\n) + n \nabla_i \left( \mu_\n +m \Phi - \frac{1}{2} m v_\n^2
   \right) + \pi_j^\n \nabla_i v_\n^j + 2\rho\epsilon_{ijk}\Omega^j v_\n^k=0  \ , \label{rot_eulern}
\ee
and
\be
\partial_t \pi_i^\s + \nabla_j (v_\s^j \pi_i^\s )
+ s \nabla_i T + \pi_j^\s \nabla_i v_\s^j  = 0 \ , \label{rot_eulers}
\ee
It is notable that the Coriolis force does not affect the
entropy equation \eqref{rot_eulers}. This is, however, not surprising. It is well-known that inertial forces are proportional to the mass, and
since our entropy component is taken to be massless it should not be affected by the Coriolis force. 
From a technical point of view, it means that the problem we consider is subtly different from the 
two-fluid r-mode problem discussed in \cite{fmode,rmode}.

Let us now consider perturbations of the rotating equilibrium models discussed in the previous section.
Considering Eulerian perturbations (denoted by $\delta$) we have, in the rotating frame (using the Cowling approximation $\delta \Phi=0$, and
not explicity denoting the velocities as perturbations since they vanish in the background
anyway);
\be
\rho \partial_t v_i^\n - 2\alpha \partial_t w_i^{\n\s} + \rho \nabla_i \delta \tilde{\mu}_\n + 2 \rho \epsilon_{ijk} \Omega^j v_\n^k = 0 \ ,
\label{comove}\ee
and
\be
2\alpha \partial_t w_i^{\n\s} + s \nabla_i \delta T = 0 \ .
\label{counter}\ee
Adding these we get an equation for the total perturbed momentum
\be
\rho \partial_t v_i^\n + \rho \nabla_i \delta \tilde{\mu}_\n + s \nabla_i \delta T + 2\rho \epsilon_{ijk} \Omega^j v_\n^k = 0 \ .
\ee
Noting that the pressure $p$ is defined by (for a co-rotating equilibrium model)
\be
\nabla_i p = \rho \nabla_i \tilde{\mu}_\n + s \nabla_i T \ ,
\ee
we see that
\be
\nabla_i \delta p = \delta \rho \nabla_i \tilde{\mu}_\n + \rho \nabla_i \delta \tilde{\mu}_\n
+ \delta s \nabla_i T + s \nabla_i \delta T \\
=  { \delta \rho \over \rho} \nabla_i p  + \rho \nabla_i \delta \tilde{\mu}_\n
+  s \nabla_i \delta T \ ,
\ee
since $\nabla_i T=0$ in the background. Hence, we have the usual Euler equation
\be
\rho \partial_t v_i^\n + \nabla_i \delta p - { \delta \rho \over \rho} \nabla_i p  + 2\rho \epsilon_{ijk} \Omega^j v_\n^k = 0 \ .
\label{comovingE}
\ee

We also have  the perturbed continuity equation for the particles
\be
\partial_t \delta \rho + \nabla_i (\rho v_\n^i) = 0 \ ,
\ee
and a conservation law for the entropy
\be
\partial_t \delta s + \nabla_i (s v_\s^i) = \partial_t \delta s + \nabla_i (s v_\n^i) - \nabla_i (s w_{\n\s}^i)  = 0 \ .
\ee
The entropy is conserved since there is no heat flux in the background ($\Gamma_\s$ is quadratic in the heat flux \cite{heat}). 

Once we provide an equation of state for matter, we  have all  relations we need to
study the linear dynamics of a non-dissipative finite temperature CFL quark core.

\subsection{Energy integrals and dissipation}

Our main aim is to work out the r-modes and establish the parameter range in
which gravitational-wave emission triggers a secular instability. In order to assess the
relevance of this instability we need to consider the various dissipative mechanisms that
counteract the growth of an unstable mode.  The damping due to shear- and bulk viscosity 
in a two-component system can be estimated using the strategy set out in \cite{fmode,rmode}. That is, we use energy integrals to 
estimate the various timescales. In essence, the damping timescale $\tau$ associated with any given process is obtained from 
\be
\tau \approx  2 E \left[ { dE \over dt} \right]^{-1}  \ ,
\ee 
where $E$ is the energy associated with the flow and $dE/dt$ is the rate of energy loss due to dissipation. This estimate 
should be accurate as long as the dissipation is weak enough that it does not affect the flow on a dynamical timescale. 

The energy associated with a given perturbation can be obtained from the equations of motion \eqref{comove} and \eqref{counter}.
Multiplying the first equation by $\bar{v}_\n^i$ and the second by $\bar{v}_\s^i$ (where the bars denote complex conjugates)
and adding, we find that the kinetic energy is given by
\be
E=\frac{1}{2}\int\rho \left( | v_\n|^2 - \frac{2\alpha}{\rho} | w_{\n\s}|^2\right) dV \ .
\ee
In the case of 
the r-modes, this is the leading order contribution to the energy.  In fact, it turns out that the second term in the bracket is 
of higher order in the slow-rotation approximation (for the same reasons as in \cite{rmode}). Hence, the r-mode energy is well approximated by 
\be
E \approx \frac{1}{2}\int\rho | v_\n|^2  dV \ .
\ee

In order to evaluate the energy loss due to viscous damping we add the dissipative part of the stress tensor $D_{ij}$ to the equations of motion. 
The total energy dissipation then follows from the body integral of 
\begin{displaymath}
- \bar{v}_\n^i \nabla^j D^\n_{ij}  - \bar{v}_\s^i \nabla^j D^\s_{ij}  \ . 
\end{displaymath}
After some algebra, this leads to
\be
{ dE \over dt} = - \int \left[ \eta \bar{\Theta}_\s^{ij} \Theta^\s_{ij} + \hat{\zeta}^{\n\n} \left| \Theta \right|^2  + 2 \hat{\zeta}^\n \mathrm{Re}\left(  \bar{\Theta} \Theta_\s \right) + \zeta \left| \Theta_\s \right|^2  \right]dV \ ,
\label{diss1}\ee 
where we have defined
\be
\Theta = \nabla_i j^i \ .
\ee
Recall the relations \eqref{exp} and \eqref{sh} and the definition $j^i = n  w_{\n\s}^i$. 

In order to combine this result with existing results for the viscosity coefficients we need to translate our variables into those 
of the ``orthodox'' two-fluid model due to, for example, Khalatnikov \cite{khalatnikov}.  
The required translation has been discussed at length in \cite{helium}, and the key relations that we need are given in
the Appendix. Carrying out the comparison, we find that the shear viscosity terms are exactly the same in the two descriptions.
Hence, we can use  $\eta$ from \cite{manuel2} without  change. In the case of the bulk viscosity, we find that the two sets of coefficients 
have different mass scalings, and we identify
\be
\zeta = \zeta_2 \quad , \quad \hat{\zeta}^\n = m \zeta_1 = m \zeta_4 \quad , \quad
\hat{\zeta}^{\n\n} = m^2 \zeta_3 \ .
\ee
This means that, when expressed in terms of the standard coefficients, the bulk viscosity dissipation rate becomes
\be
{ dE \over dt} =
 - \int \left[ \eta \bar{\Theta}_\s^{ij} \Theta^\s_{ij} + m^2 \zeta_3 \left| \Theta \right|^2  +  2 m\zeta_1 \mathrm{Re}\left(  \bar{\Theta} \Theta_\s \right) + \zeta_2 \left| \Theta_\s \right|^2  \right]dV \ .
\label{diss2}\ee 

To conclude,  there are  three bulk viscosity  coefficients rather than the usual single one, and both dynamical degrees of freedom are needed if we want to evaluate the dissipation integrals. 
Note that the dissipation is expressed in terms of $j^i$ and $v_s^i$, rather than $v_\n^i $ and $w_{\n\s}^i$, the variables that we will solve for when we determine the r-modes. 
It is, of course, straightforward to express \eqref{diss2} in terms of these variables.  Unfortunately, the  dissipation integrand is then rather messy. We find  
\begin{multline}
{ dE \over dt} =  - \int \Big\{  \eta \left[ \bar{\Theta}^{ij}_\n \Theta_{ij}^\n - 2 \mathrm{Re} \left( \bar{\Theta}^{ij}_\n \Theta_{ij}^{\n\s} \right) + \bar{\Theta}^{ij}_{\n\s} \Theta_{ij}^{\n\s} \right] \\
+ \zeta_2 \left| \Theta_{\n} \right|^2 + \left[ \zeta_2 - 2 \rho \zeta_1 + \rho^2 \zeta_3 \right] \left| \Theta_{\n\s} \right|^2  - 2 ( \zeta_2 - \rho \zeta_1) \mathrm{Re} \left( \bar{\Theta}_\n \Theta_{\n\s} \right) \\ + 2 \zeta_1  \mathrm{Re} \left[ \bar{\Theta}_\n \left( w_{\n\s}^j \nabla_j \rho \right) \right]
- 2 ( \zeta_1 - \rho \zeta_3)  \left[ \bar{\Theta}_{\n\s} \left( w_{\n\s}^j \nabla_j \rho \right) \right]
+ \zeta_3 \left|   w_{\n\s}^j \nabla_j \rho \right|^2  \Big\} dV \ .
\label{complications}\end{multline}
where
\be
\Theta_\n = \nabla_j v_\n^j \quad \mbox{ and } \quad 
\Theta_{\n\s} = \nabla_j w_{\n\s}^j \ .
\ee
However, as we will see later, this expression simplifies considerably in the r-mode problem. 

Before moving on, it is worth making a general point. It might be tempting to suggest that the terms involving $w_{\n\s}^i$ in \eqref{complications} should be 
less relevant than those involving only $v_\n^i$. However, without solving for an actual oscillation mode one cannot make this argument precise. 
In general, the  bulk viscosity contributions from  terms involving $w_{\n\s}^i$ cannot be neglected. Since a mode-oscillation typically involves both degrees of freedom
(unless the star is unstratified 
\cite{prix2,passamonti}), we need to determine the nature of the oscillations before we make  further simplifications.

\section{A simple model equation of state}

In order to obtain quantitative results for the r-mode instability we need to provide an equation of state.
As our main focus is on the two-fluid aspects of the problem (ignored in previous studies), 
the equation of state must account for the thermal excitations, i.e. the phonons.
Moreover, if we want to investigate the relative importance of the different bulk viscosity terms we need to allow for a relative flow at the perturbative level.
This is also important if we want to consider the superfluid mutual friction \cite{mana1}. However, if we  insist on the model being truly ``realistic'' then
the problem becomes much more challenging. Hence, we will make a number of simplifying assumptions. This is not only practical, it is also quite reasonable
since this is our first attempt at a quantitative analysis. Future work should  aim to relax some of our
assumptions.

As discussed in Section~II, the equation of state that we require takes the form of an energy functional $E=E(n,s,w^2)$.
Once this functional is provided, we can work out all the quantities that we need  to construct a rotating model and
 calculate the r-modes. Unfortunately, we do not have an equation of state of the required form. In fact, some of the parameters
that we need are usually not considered, essentially since they are not required if one is only interested in equilibrium configurations.
In order to make progress we have to build a suitably simple model equation of state. To do this, we assume that the thermal contribution due to the phonons
 adds to a zero-temperature equation of state for the condensate. This strategy has previously been developed for a relativistic superfluid \cite{cl95}, 
and builds on the classic expressions for a non-relativistic phonon gas, see \cite{khalatnikov}.

\subsection{The phonon gas}

Let us assume that, at zero temperature matter is described by a barotropic model such that $E(n,s=0)=E_0(n)$. Then the chemical potential
follows from
\be
\mu = { d E_0 \over d n} \ ,
\ee
the pressure is obtained from
\be
dp_0 = n d \mu \ ,
\ee
and it is easy to work out the speed of (first) sound;
\be
c_0^2 = { dp_0 \over d\rho } \ ,
\ee

Now consider thermal excitations represented by a phonon gas with a linear dispersion relation,
with slope $c_0$. Working out the thermodynamics of such a gas one can show that its contribution
to the pressure, $\psi$, is given by \cite{cl95}
\be
\psi = { 4 \pi^5 \over 45} { (kT)^4 \over (2 \pi \hbar c_0)^3} \left( 1 - {w^2 \over c_0^2} \right)^{-2} \ .
\label{psidef}\ee
Here, and in the following, it is to be understood that $w^2$ corresponds to $w^2_{\n\s}$.
From this we obtain the entropy density (in the matter frame) via 
\be
s = { \partial \psi \over \partial T} \ .
\ee
This leads to the explicit result;
\be
s = {4 \psi \over kT} = { 16 \pi^5 \over 45}  \left( {kT \over 2 \pi \hbar c_0} \right)^3
\left( 1 - {w^2 \over c_0^2} \right)^{-2} \ ,
\label{srel}\ee
which means that the heat capacity is given by
\be
c_\mathrm{v} = T { \partial s \over \partial T } = { 2 \pi^2 \over 15} \left( {kT \over \hbar c_0} \right)^3 \ ,
\ee
in  agreement with the phonon result in \cite{alford2}.

Finally, one can show that \cite{cl95} the ``normal fluid'' density, $\rho_\N$, required in the orthodox
superfluid formalism (see Appendix) is given by
\be
\rho_\N = { 16 \pi^5 \over 45} {1 \over (2 \pi \hbar)^3} \left( {kT \over c_0} \right)^4 \left( 1 - {w^2 \over c_0^2} \right)^{-3} {1 \over c_0} \ .
\ee
This result is consistent with
\be
\rho_\N = { 4 \over 3} { E_\mathrm{ph} \over c_0^2} \left( 1 - {w^2 \over c_0^2} \right)^{-3} \ .
\ee
where the phonon energy is
\be
E_\mathrm{ph} = { 4 \pi^5 \over 15} \left( {kT \over 2 \pi \hbar c_0} \right)^3 kT \ .
\ee
In our formulation of the problem, we need the entrainment between particles and  entropy.
As discussed in \cite{helium} the relevant entrainment parameter is related to $\rho_\N$ according to
\be
\alpha = - { \rho_\N \over 2} \left( 1 - { \rho_\N \over \rho} \right)^{-1} \approx - { \rho_\N \over 2} \ ,
\label{entrain}\ee
where the last approximation is accurate in the low temperature limit, when $\rho_\N \ll \rho$.
This is the regime that we are considering here.

It is worth noting that the above results lead to
\be
\rho_\N \approx 29 \left( {T \over 10^9 \mathrm{K}} \right)^4 \mathrm {g/cm}^3 \ .
\ee
In order to be consistent we need $\rho_\N \ll \rho \approx 7\times 10^{14} \mathrm{g/cm}^3$ (the average density for a canonical neutron star),
which translates into $T\ll 2 \times 10^{12}\ \mathrm{K}$. In other words, the
model should be valid for all astrophysical neutron stars.

Let us now use the phonon gas results to construct a ``complete'' model equation of state.
That is, we want to deduce a consistent energy functional $E$. This functional should be such that
the temperature is obtained from 
\be
T = \left. { \partial E \over \partial s} \right|_{n,w^2} \ .
\ee
Now, combining \eqref{psidef}  with \eqref{srel}  we find that
\be
\left. { \partial E \over \partial s} \right|_{n,w^2} = { 1 \over B} s^{1/3} \ ,
\ee
where
\be
B = \left( {16 \pi^5 \over 45} \right)^{1/3} { 1 \over 2 \pi \hbar c_0 }  \left( 1 - {w^2 \over c_0^2} \right)^{-2/3} \ .
\ee
Since $c_0=c_0(n)$ (by definition) we can integrate to get
\be
E= E_1(n,w^2) + { 3 \over 4B} s^{4/3} = E_1 + 3\psi = E_1 + E_\mathrm{ph} \left( 1 - {w^2 \over c_0^2} \right)^{-2} \ .
\ee
We learn that in the limit of a low relative velocity ($w\ll c_0$) the equation of state is simply given
by
\be
E = E_1 + E_\mathrm{ph} \ .
\ee
This result is quite intuitive. Using the fundamental relation of thermodynamics
\be
p + E = n\mu + sT \ ,
\ee
it is also straightforward to show that the total pressure is given by
\be
p = p_0 + \psi \ ,
\ee
as one might have expected.

Now that we have the required energy functional, we can determine the entrainment parameter from
\be
\alpha = \left. { \partial E \over \partial w^2 } \right|_{n,s} \ .
\ee
Working this out, we find that (after some algebra)
\be
\alpha = {\partial E_1 \over \partial w^2} - { \rho_\N \over 2} \ .
\ee
Comparing to \eqref{entrain} we see that the model is consistent with
\be
{\partial E_1 \over \partial w^2} = 0 \ .
\ee
Hence, it is natural to take $E_1=E_0(n)$, i.e.
simply add the thermal phonon energy to the zero-temperature equation of state.
The final result is then
\be
E= E_0 + 3\psi = E_0 + E_\mathrm{ph} \left( 1 - {w^2 \over c_0^2} \right)^{-2} \ .
\ee

In order to proceed, we need to provide the zero-temperature equation of state. A ``realistic'' model
should obviously be based on QCD. It is natural \cite{jaikumar} to use the MIT bag model.
In the case of CFL matter, this  leads to (using units where the speed of light is unity, $c=1$)
\be
p_0\approx { 3 \mu^4 \over 4 \pi^2} - B + {3\mu^2 \over 4\pi^2} (4 \Delta^2 - m_s^2) \ ,
\label{p0}\ee
and 
\be
\rho \approx { 9 \mu^4 \over 4 \pi^2 } + B - { 3 \mu^2 m_s^2 \over 4\pi^2} \ ,
\ee
where $B$ is the bag constant and $\Delta$ is the pairing gap associated with the CFL condensate.

For simplicity, we will use 
\be
p_0 \approx { 1 \over 3} ( \rho  -4B) \ ,
\label{bag}\ee
in which case the speed of sound is constant;
\be
c_0^2 = { dp_0 \over d \rho} \approx { 1 \over 3} \ .
\ee
As discussed in \cite{jaikumar,alford3}, this approximation may be quite good, essentially because the two contributions to the last term in \eqref{p0} 
almost exactly cancel each other. Later, when we work out the various r-mode dissipation integrals, we will simplify the problem further 
by assuming that the density is uniform, with a constant speed of sound.

\subsection{Viscosity}

In order to determine the damping timescale for the unstable r-modes, we need to supplement the
non-dissipative model with viscous terms. From the discussion in section IIIB we know  that we need one shear viscosity coefficient and
three bulk viscosity coefficients. In order to account for dissipation associated with the
superfluid vortices, we should also consider the mutual friction. Some of the required viscosity coefficients
have been discussed in the literature. In particular, the dissipation due to various phonon interactions
has been investigated \cite{manuel1,manuel2,mana1}. In fact, the availability of these results 
is one of the key reasons for us focussing on the condensate plus phonon model. Having said that, the available
 results are  incomplete. To what extent this is the case, and how we have dealt with
this problem, will be discussed below. 

Before we proceed it is worth iterating the point that we are focusing on the phonons because they 
represent the simplest in a hierarchy of relevant multi-fluid problems. The presence of kaons, either as thermal excitations or
a condensate \cite{alford1,alford2}, would require us to extend the model to account for additional degrees of freedom. 
The issues involved are similar to those for  a superfluid hyperon core \cite{hyperon}, and we expect to
consider them in the  future. 

Let us first consider the phonon shear viscosity. The relevant viscosity 
coefficient is determined in  \cite{manuel2}. We should use (following the discussion above, we are using $c_0^2=1/3$ here)
\be
\eta = 2.46 \times 10^{26} \left( { \mu_q \over 300\ \mathrm{MeV} } \right)^8 \left( { 10^9 \mathrm{K} \over T} \right)^5  \ \mathrm{g/cm\ s} \ ,
\label{eta0}\ee
where $\mu_q$ is the quark chemical potential. We arrive at this result by taking both  $\mu_\n$ and the 
quark ``mass'' $m$ in the fluid model to be equal to the quark chemical potential $\mu_q$.
The temperature scaling of $\eta$ is the same as in the case of $^4$He, which makes sense given that the involved phonon
processes are the same. However, in the context of neutron stars, this model is severely limited.
Basically, the ``hydrodynamic'' treatment of the phonons is no longer valid if their mean-free path
is larger than (or comparable to) the size of the system. As in kinetic theory, the mean-free
path associated with the shear viscosity follows from (up to a factor of order unity)
\be
\eta \simeq \rho_\N c_0 \lambda \ .
\ee
For the phonon gas model this leads to
\be
\lambda \simeq 5 \times 10^{14} \left( { \mu_q \over 300~\mathrm{MeV} } \right)^8 \left( { 10^9 \mathrm{K} \over T} \right)^9 \ \mathrm{cm} \ .
\ee
In other words, for a 10~km star the model will not be valid below $\sim 10^{10}~K$. This is obviously a problem, since mature neutron stars 
are expected to be significantly colder than this. In essence, we ought to model the phonons as ballistic in the low temperature regime. 
However, there is some evidence from laboratory experiments on $^4$He that the fluid model remains relatively accurate also at lower temperatures.
Given this, we will use the two-fluid model also in the regime where it is no longer formally valid. 
Still, in order to describe such systems we need a different model for the shear viscosity. 
In order to make progress we will simply assume that the
mean free path is limited by the size of the system. That is, we will use an \underline{effective} shear viscosity
given by
\be
\eta_\mathrm{eff} \simeq  \rho_\N c_0 R \ ,
\ee
where $R$ is the radius of the star. This may seem like a rather drastic assumption, but it has recently been shown \cite{helsound}
that in the case of Helium it leads to results that agree quite well with the observed sound attenuation.
In particular, we would have  $\eta_\mathrm{eff} \sim \rho_\N \sim T^4$  at low temperatures, which means that the viscosity weakens
as the phonon
density decreases. This makes (at least qualitative) sense. Explicitly, we get
\be
\eta_\mathrm{eff} \approx 5 \times 10^{17} \left( {T \over 10^9\ \mathrm{K}} \right)^4\ \mathrm{g/cm s} \ .
\label{etaeff}\ee
This model produces a maximum in the
viscosity as a function of temperature, a feature that seems quite natural. Having said that, more detailed work on the low-temperature phonon problem is
obviously needed in order to 
improve on our results. This is, in fact, a very interesting problem; when the mean-free path is large, the phonons interact with the ``surface''
more frequently than with each other. In principle, the hydrodynamics approach should not be valid in this regime. Yet, there is some evidence
from studies of heat conduction in nano-systems (see \cite{heat} for a recent discussion) that a judicious choice of surface boundary condition for the 
phonons leads to a useful ``fluid'' model. We plan to explore this idea further in the future.

We now turn to the bulk viscosity. Only one of the three required coefficients has, so far, been calculated in detail.
In the static limit, the analysis in \cite{manuel1} leads to the result
\be
\zeta_2(\omega=0) = 1.75 \times 10^{12} \left( {m_s \over 100 \mathrm{MeV} }\right)^4 \left( {10^9~\mathrm{K} \over T} \right) \ \mathrm{g/cm s} \ ,
\label{zeta0}
\ee
where $m_s$ is the strange quark mass.
However, in  order to use this result to study oscillation modes, we  need consider the frequency dependence in more detail. 
This is essential since the bulk viscosity is a ``resonant'' mechanism that is particularly effective when the involved dynamics has a timescale similar to the 
relevant relaxation time. The static limit  only provides partial information.  
Following \cite{mana2} we will use
\be
\zeta_i = \frac{\tau}{1+\omega^2\tau^2}\alpha_i(T) \ , \qquad i = 1-3 \ ,
\label{zeds}
\ee
where the ``amplitudes'' $\alpha_i$, in general, depend on the different physical scales of the system, 
while $\tau$ is the relaxation time for the processes that give rise to bulk viscosity. According to \cite{mana2}, the relaxation time
scales according to
\be
\tau \sim \frac{c_0^3\mu_q^8}{T^9} \ .
\label{relax}
\ee
The scaling  of all three viscosity coefficients, in the static limit, has been determined in \cite{mana2}. We should have
\beq
\zeta_1&\approx&\tilde{\alpha}_1\frac{m_s^2}{T\mu_q} \ , \\
\zeta_2&\approx&\tilde{\alpha}_2 \frac{m_s^4}{T}  \ ,\\
\zeta_3&\approx&\tilde{\alpha}_3 \frac{1}{T\mu_q^2} \ ,
\eeq
where $\tilde{\alpha}_i$ are constants. This, together with the relations in \eqref{zeds} and \eqref{relax}  allows us to determine the frequency dependent bulk viscosity coefficients, up to the unknown amplitudes $\tilde{\alpha}_i$ and $\beta$. These can not be determined with the approximations used in \cite{mana2}, although we can obviously infer $\tilde{\alpha}_2$ from \eqref{zeta0}.
From the analogous problem for $^4$He, we know that the three bulk viscosity coefficients may be of the same order of magnitude.
In absence of detailed results for the quark case, it thus makes sense to assume that  $\zeta_1$ and $\zeta_3$ have a similar amplitude to $\zeta_2$. 
In particular, this would suggest that we parametrise  $\zeta_3$, in the static limit, according to
\be
\zeta_3(\omega=0) {\left({\rho \over {10^{14}~\mathrm{g/cm}^3 }} \right)}^{2} = \bar{\alpha} \times 10^{12} \left( {10^9~\mathrm{K} \over T} \right) \left( {300~\mathrm{MeV} \over \mu_q} \right) \ \mathrm{g/cm s} \ ,
\label{zeta30}
\ee
where $\bar{\alpha}$ is an unspecified parameter.
The frequency dependent result then takes the form
\be
\zeta_3=\frac{\zeta_3(\omega=0) }{1+\omega^2\tau^2}\label{z3} \ ,
\ee
with
\be
\tau=\beta\times 10^7 \left( {10^9~\mathrm{K} \over T} \right)^9 \left( {\mu_q \over 300~\mathrm{MeV} } \right)^8 s\label{ts} \ .
\ee
Here, $\beta$ is another undetermined parameter. In principle, one would expect both $\bar{\alpha}$ and $\beta$  to be of order unity, but given the lack of precise information we
can vary them and assess how this affects the bulk viscosity damping of the r-modes. 
Note that the coefficient in \eqref{z3} has a maximum at the resonance frequency $\omega=1/\tau$. We know from previous work that it is essential to understand 
how this resonance frequency relates to the  r-mode frequency. A precise statement to this effect is not possible, given the free parameters in our model, 
but the strong temperature dependence in \eqref{ts} means that even a change of several orders of magnitude in $\tau$ would produce a relatively small 
change in the temperature at which the resonance occurs for a given mode frequency. This suggests that the uncertainty associated with $\beta$ may not 
affect our analysis too severely (assuming that the temperature scaling is correct, of course).
If we, for example, consider a mode frequency $\omega=10^4$ s$^{-1}$ and an amplitude such that $\tau=10^8$~s, we would have a resonance around $T\approx 3.5\times 10^8$ K.
If, on the other hand,  we take a drastically shorter relaxation timescale of $\tau=1$ s, then the resonance appears near $T\approx 2.8\times 10^9$ K. That is, 
the resonance temperature would shift by less than an order of magnitude.
This model is obviously phenomenological, and allows us to proceed, but
an actual derivation of the frequency dependent viscosity coefficients is needed if we want more detailed results. 

It is obviously  important to compare our results to other relevant estimates for the r-modes. The natural, and most immediate,
comparison would be to  unpaired quark matter. In that case, which corresponds to a single fluid problem, 
we have \cite{jaikumar,madsen}
\be
\eta \approx 1.4 \times10^{17} \left( {0.1 \over \alpha_s} \right) \left({\rho \over 10^{14} \mbox{g/cm}^3}\right)^{5/3} 
\left( { T \over 10^9\ \mathrm{K} } \right)^{-2} \ \mathrm{g/cm s} \ ,
\label{unpeta}\ee
where $\alpha_s$ is the strong interaction coupling constant, 
and
\be
\zeta = { A(T) \over \omega^2 + B(T)}  \ ,
\label{unpzeta}\ee
where
\be
A(T) = 1.8\times10^{36} \left( {\mu_q \over 300 \mathrm{MeV} }\right)^3 \left( {m_s \over 100 \mathrm{MeV}} \right)^4
\left( { T \over 10^9\ \mathrm{K} } \right)^2 \ \mathrm{g/cm s}^3 \ ,
\ee
and
\be
B(T) = 2.6 \times 10^{7} \left( {\mu_q \over 300 \mathrm{MeV} }\right)^6 \left( 1 + {m_s^2 \over 4 \mu_q^2} \right)^2
\left( { T \over 10^9\ \mathrm{K} } \right)^4 \ \mathrm{s}^{-2} \ .
\ee

We also want to quantify the relevance of the vortex mutual friction for the r-mode instability in CFL matter. In order to do this we need a
representation of the counter-moving degree of freedom associated with an r-mode (which is of order $\Omega^2$ in the slow-rotation approximation) \cite{rmode}. 
Before we consider this problem, we need to translate the mutual friction results from \cite{mana1} 
to our formalism. In order to relate the parameter $\mathcal{B}$ to the results in \cite{mana1}, let us consider equation \eqref{counter} where we now include a mutual friction 
force of the form \eqref{mf}. This leads to an equation for the evolution of the relative velocity of  form
\be
\frac{\partial w_i^{\n\s}}{\partial t}+\frac{s}{2\alpha}\nabla_i \delta T=- { \rho \over 2 \alpha} \left[ {\mathcal{B}'}  n_v \epsilon_{ijk} \kappa^j w_{\n\s}^k
+ \mathcal{B}  n_v \epsilon_{ijk} \epsilon^{klm} \hat{\kappa}^j 
\kappa_l w_m^{\n\s}\right] \ .
\ee
Given this equation, and the results in the Appendix, we can compare our formalism to the standard mutual friction description, see e.g. equations (2.2)-(2.3) in \cite{book}.
This allows us to determine $\mathcal{B}$ from the parameter $\tilde{\alpha}$  that is calculated in \cite{mana1}. This leads to 
\be
\mathcal{B} = { 2 \pi^5 \over 405 } \left(1-{ 2 \alpha \over \rho} \right){\left(1-c_0^2 \right)^2 \over c_0^6} \left({T \over \mu_q}\right)^5 \ ,
\ee
or
\be
\mathcal{B} \approx 3.5\times 10^{-17}\left(1-{ 2 \alpha \over \rho} \right) \left( {300\ \mathrm{MeV} \over \mu_q} \right)^5 \left( {T \over 10^9\ \mathrm{K}} \right)^5 \ .
\label{Bcoff}\ee
We  see that the mutual friction vanishes as the temperature decreases. This obviously makes sense
since there will be no phonon-vortex scattering when the phonon gas becomes dilute. Moreover, according to this result, we are safely in the extreme ``weak drag'' regime (c.f. the discussion in \cite{prec}) where
\be
\mathcal{B}' \approx \mathcal{B}^2 \ll 1 \ .
\ee
This implies that the phonon mutual friction will not damp the r-modes efficiently (probably as expected) \cite{rmode}. 
This result is confirmed by the detailed analysis in section~\ref{mutualf}. However, these results come with an important caveat. 
Strictly speaking, our analysis is only valid as long as the phonons can be described as a fluid. As we have already explained, this is not 
the case for mature neutron stars. In fact, the results of \cite{mana1} are supposedly derived for ballistic phonons. This means that our analysis is
somewhat inconsistent. At temperatures above (say) $10^{10}$~K the mutual friction parameter from \cite{mana1} may not apply, and below this temperature
our two-fluid model may not be appropriate. However,  the latter issue may not be that important. After all, one would expect the phonon
mutual friction to weaken dramatically at lower temperature, c.f. \eqref{Bcoff}, meaning the any quantitative errors in the analysis will be of no 
real importance. 

\section{Approximate r-mode results, shear- and bulk viscosity}

Our (simple) model for a two-component cool CFL core is now ``complete'', and 
we may conduct the r-mode analysis as in \cite{fmode,rmode}. 
In order to quantify the damping due to bulk viscosity, we need to account for terms of order $\Omega^2$ in the analysis. This complicates the problem,
since the centrifugal deformation of the star's shape enters at the same level. In view of this, and the fact that this is a first exploratory study, we
will make a sequence of approximations that allow us to proceed analytically. This strategy also makes sense since we are using a simplified equation of state.
The results of our analysis should help determine whether 
a full numerical study is worthwhile. 

First of all, we need to make a choice of primary variables. The linearised equations link the four scalar variables $\delta p$, $\delta T$, $\delta s$ and $\delta \rho$.
In principle, we can use the equation of state to express any two of these in terms of the other two. Following the strategy set out in \cite{rmode} we will
work with  $\delta p$ and $\delta T$. For our model equation of state we have (since the background is co-rotating)
\be
\delta p = \delta p_0 + \left( {\partial \psi \over \partial T} \right) \delta T = \delta p_0 + s \delta T \ ,
\ee
which means that we can express the equations in terms of $\delta p_0$ and $\delta T$.
We then have
\be
\delta \rho = \left( {\partial \rho \over \partial p_0} \right) \delta p_0 +  \left( {\partial \rho \over \partial T} \right) \delta T = { 1 \over c_0^2} \delta p_0 \ ,
\ee
since $c_0$ is contant, and
\be
\delta s =  \left( {\partial s \over \partial p_0} \right) \delta p_0 +  \left( {\partial s \over \partial T} \right) \delta T =   { c_\mathrm{v} \over T} \delta T \ ,
\label{ds}\ee
since $p_0=p_0(n)$.

In order to avoid numerics we will make use of a ``trick'' used by
Lindblom, Owen and Morsink in one of the early r-mode instability papers \cite{lom}. The basic idea is to neglect the rotational
change in shape in different terms in the perturbation equations. Once this is done one can
``estimate'' the bulk viscosity  from the leading order r-mode solution. Although this simplification is not 
not based on a rigorous argument, one can show that the bulk viscosity damping timescale is of the right order of magnitude
(within about a factor of 5 of the true second-order slow-rotation result \cite{lmo}). This is good enough for our
present purposes.  

In our case, the calculation
would proceed as follows: First consider the continuity equation
\be
i \omega \delta \rho + \rho \nabla_i v_\n^i + v_\n^i \nabla_i \rho = 0 \ ,
\ee
where we have assumed that all perturbed quantities behave as $e^{i(\omega t + m\varphi)}$.
If we simply omit the last term (that would vanish for a spherical star since an r-mode is purely toroidal to leading order), we get
\be
\nabla_i v_\n^i \approx -{ i \omega \over \rho} \delta \rho = - { i \omega \over \rho c_0^2} \delta p_0 \ .
\ee
This may not be a very accurate estimate, but
it allows us to progress without having to determine the rotational corrections to the r-mode (which depend on 
the term that we have neglected).
Now consider the second degree of freedom in a similar way. From the entropy equation we get
\be
i \omega \delta s + s \left( \nabla_i v_n^i - \nabla_i w_{\n\s}^i \right) + (v_\n^i - w_{\n\s}^i ) \nabla_i s = 0\ .
\label{dels}\ee
In fact, for the model equation of state the last term vanishes identically since $s=s(T)$ and $\nabla_i T =0$ for our background model.
It follows that
\be
 \nabla_i v_\n^i - \nabla_i w_{\n\s}^i \approx -i \omega { \delta s \over s}  \ .
\ee

The approximation is completed by the $\varphi$-components of the two Euler equations. 
Assuming that the r-mode velocity field takes the form
\be
v_\n^i = \left[  v_r \mathbf{e}_r^i+ v_\theta \mathbf{e}_\theta^i + v_\varphi \mathbf{e}_\varphi^i \right]  e^{i(\omega t + m\varphi)} \ ,
\label{vector}\ee
and that  $w_{\n\s}^i$ is similar, although of higher slow-rotation order than $v_\n^i$ \cite{rmode}, we get
\be
i\omega \rho  r^2 \sin^2 \theta v_\varphi + im \delta p + 2 \rho \Omega r^2 \sin \theta \cos \theta v_\theta = 0 \ ,
\ee
and
\be
2i\alpha \omega   r^2 \sin^2 \theta w_\varphi + ims \delta T = 0 \ .
\label{dTeq}\ee
The second relation shows immediately that $\delta T$ is of higher slow-rotation order for
the classic r-mode (since $w_\varphi$ is of higher order and $\omega \sim \Omega$ for inertial modes). This means that we can justifiably  neglect
the temperature variation in the pressure term in the first equation. That is, we have
\be
i\omega \rho  r^2 \sin^2 \theta v_\varphi + im \delta p_0 + 2 \rho \Omega r^2 \sin \theta \cos \theta v_\theta = 0 \ .
\label{dpeq}\ee
This is exactly the result used in \cite{lom}. Moreover, we now see that the 
first term in \eqref{dels} should also be of higher order, which means that we are left with
\be
\nabla_i w_{\n\s}^i \approx  \nabla_i v_\n^i  \ .
\label{compress}\ee

The above relations allow us to estimate the terms needed to evaluate the bulk viscosity 
integrals once we have the leading order  r-mode solution. 
Since the required contribution to the r-mode velocity field is purely toroidal, we have
\be
v_\theta =  - { U_l \over r^2 \sin \theta} \partial_\varphi Y_l^m \ , \qquad v_\varphi =   { U_l \over r^2 \sin \theta} \partial_\theta Y_l^m  \ .
\ee
Moreover, the mode-solution is such that the only contribution comes from  $l=m$ in which case
 $U_m=(r/R)^{m+1}$ (noting that the normalisation is irrelevant at the linear perturbation level). Finally, to the accuracy needed, the 
mode frequency is
(in the rotating frame)
\be
\omega = { 2\Omega \over m+1} \ .
\label{rfreq}\ee

Using these results, and expanding the (scalar) pressure perturbation in spherical harmonics, i.e. using $\delta p_0 = \sum_l \delta p_l Y_l^m e^{i(\omega t + m\varphi)}$, we find that 
\eqref{dpeq} leads to 
\be
{ \delta p_{m+1} \over \rho} = { 1 \over \sqrt{2m+3}} { 2  m \Omega \over m+1}  U_m \ ,
\ee
and thus we have
\be
\nabla_i v_\n^i \approx \nabla_i w_{\n\s}^i \approx - { i \over \sqrt{2m+3}} { 4 m \Omega^2 \over (m+1)^2 c_0^2} U_m Y_{m+1}^m \ .
\label{dv}\ee
This completes the approximation scheme, which provides all the expressions we need to make a rough estimate of the bulk viscosity
damping timescale.

\section{Mutual friction}
\label{mutualf}

In order to estimate the mutual friction damping, we need to determine the detailed counter-moving degree of freedom 
associated with the r-modes. In principle, this calculation proceeds as in \cite{rmode} and involves the rotational corrections to the shape of the star. 
However, in order to be consistent we will instead build on the approximations introduced in the previous section. 
First of all, it is easy to show that the countermoving degree of freedom will be poloidal (in contrast to the toroidal 
leading order r-mode solution). This means that we have
\be
w_r = {1\over r} W_l Y_l^m \ , \qquad w_\theta =   { V_l \over r^2} \partial_\theta Y_l^m \ , \qquad w_\varphi =   { V_l \over r^2 \sin \theta} \partial_\varphi Y_l^m  \ .
\ee
If we also expand the temperature perturbation in such a way that $\delta T = \sum_l T_l Y_l^m$, then it follows from \eqref{dTeq} that
\be
 V_l = { i s \over 2 \alpha\omega } T_l  \ .
\ee 
Similarly, we get from the radial component of \eqref{counter};
\be
 W_l = { i s \over 2 \alpha\omega }\partial_r T_l  \ .
\ee 
Combining these results with \eqref{rfreq} and \eqref{dv}, we find that
\be
s T_{m+1} = { 8m\alpha \Omega^3 \over \sqrt{2m+3} (m+1)^3 (2m+5)} {R^2 \over c_0^2} \left( {r \over R} \right)^{m+1} \left[ {m+3 \over m+1} - 
\left( {r \over R} \right)^2 \right] \ .
\label{Tsol} \ee
From this result we easily obtain the required counter-moving components of the  r-mode. 

We arrive at \eqref{Tsol} by assuming that $w_r=0$ at the surface (in accordance with the discussion in \cite{rmode}). This condition is unlikely to be  correct
in the present problem, and a detailed analysis of the appropriate condition to impose on the phonons needs to be carried out in the future (c.f. the  
discussion in \cite{heat}). At this point we 
simply note that our assumption that the whole star is composed of CFL matter is artificial anyway. Moreover, the boundary condition 
does not affect the parameter scaling of the result. This is the dominant factor in determining the r-mode damping timescale.  

\section{The r-mode instability window}

To estimate the growth/damping timescales for the r-mode instability
we need to evaluate  various energy integrals, see \cite{lom,review}.  First of all, the mode energy is given by 
\be
E = { 1 \over 2} m(m+1) R^{-(2m+2)} \int_0^R \rho r^{2m+2} dr \ ,
\ee
while the
growth time-scale due to (current multipole) gravitational-wave emission is obtained from
\be
{ 1 \over \tau_\mathrm{gw} } = - { 32 \pi G \Omega^{ 2m+2} \over c^{2m+3}}
{(m-1)^{2m}\over [ (2m+1)!!]^2} \left( {m+2 \over m+1} \right)^{2m+2} \int_0^R \rho r^{2m+2} dr \ .
\ee
Turning to the shear viscosity, we know that $|w_{\n\s}|\ll |v_\n|$ for the superfluid r-mode. This means that the shear viscosity integral 
simplifies to the usual, single fluid result. Hence, the leading-order energy loss due to shear viscosity follows from the  
expression 
\begin{multline}
\dot{E}_\mathrm{sv}=- m(m+1)\left\{\int_0^R { \eta \over  r^2} \left[ r^6 \left| \partial_r \left({ U_m \over r^2} \right) \right|^2+(m-1)(m+2) |U_m|^2\right] \right\} dr \\
= - m(m^2-1)(2m+1) R^{-(2m+2)}  \int_0^R \eta r^{2m} dr  \ ,
\label{1shear}
\end{multline}
leading to
\be
{ 1 \over \tau_\mathrm{sv}} = - (m-1)(2m+1)  \int_0^R \eta r^{2m} dr \left[ \int_0^R \rho r^{2m+2} dr \right]^{-1} \ .
\ee

Next, we need to estimate the bulk viscosity damping. From \eqref{complications} we know that the general bulk viscosity expression is complicated, especially since we cannot rule out the possibility that
the three different bulk viscosity  contributions are of similar magnitude. However, the approximations that we have made simplify the problem considerably. Neglecting the terms
that arise due to the change in shape of the rotating star [the last line in \eqref{complications}], and making use of
\eqref{compress} we find that 
\be
\dot{E}_\mathrm{bv} \approx - \int_0^R \zeta_\mathrm{eff} \left| \nabla_i v_\n^i \right|^2 dV
\approx - { 16 m^2 \Omega^4 \over (m+1)^4 (2m+3) }  R^{-(2m+2)}  \int_0^R { \zeta_\mathrm{eff} \over c_0^4} r^{2m+4} dr \ ,
\ee
which leads to 
\be
{ 1 \over \tau_\mathrm{bv}} = - { 16 m \Omega^4 \over (m+1)^5 (2m+3) }  \int_0^R { \zeta_\mathrm{eff} \over c_0^4} r^{2m+4} dr \left[ \int_0^R \rho r^{2m+2} dr \right]^{-1} \ ,
\ee
where
\be
\zeta_\mathrm{eff} \approx  \rho^2 \zeta_3 \ .
\ee
This result follows after some surprising cancellations, associated with \eqref{compress}. It is notable that only one of the three 
bulk viscosity coefficients plays a role in our simplified case. Moreover, it is \underline{not} the one ($\zeta_2$) 
that remains in the Navier-Stokes limit. Of course, one would not expect this drastic simplification 
for a more realistic stellar model. This fact provides strong motivation for a more detailed, numerical, r-mode analysis.
We also learn that we can not, in general, ignore the additional bulk viscosities in the multi-fluid problem. 

Finally, we have the dissipation integral for the mutual friction. From the results discussed in \cite{fmode,rmode}, we find that
\be
\dot{E}_\mathrm{mf} \approx - 2 \int \rho \mathcal{B} \Omega \left( \delta_i^m - \hat{\Omega}^m \hat{\Omega}_i \right) w_{\n\s}^i \bar{w}^{\n\s}_m dV \ .
\ee
 Working out the angular integrals, this reduces to
\be
\dot{E}_\mathrm{mf} \approx -2 \int \rho \mathcal{B} \Omega \left\{ \left[ 1- Q_{m+2}^2 - Q_{m+1}^2 \right] W_{m+1}^2 +
\left[ (m+1)(m+2) - (m+1)^2 Q_{m+2}^2 - (m+2)^2 Q_{m+1}^2 \right] V_{m+1}^2 \right\} dr \ ,
\label{mfint}\ee
where we have used 
\be
Q_l^2 = { (l+m)(l-m) \over (2l-1)(2l+1)} \ .
\ee
The integral in \eqref{mfint} is easily evaluated using the eigenfunctions from the previous section. 

In order to evaluate the relevant integrals, and obtain estimated timescales with explicit parameter scaling, we now assume that the 
density is uniform. Taking all parameters constant, we easily find that
\be
\tau_\mathrm{gw} = -22 \left( { 1.4 M_\odot \over M} \right) \left( { 10\ \mathrm{km} \over R} \right)^4 \left( { P \over 1\ \mathrm{ms}} \right)^6 \ \mathrm{s} \ ,
\label{tgw}\ee
for the $l=m=2$ r-mode (the sign indicates that the mode is unstable).
We also get, for the shear viscosity,
\be
{1 \over \tau_\mathrm{sv}} \approx  1.1 \times 10^{-26}\left( { \eta \over 1\ \mathrm{g/cms} } \right) \left( { R\over 10\ \mathrm{km} } \right)\left( { 1.4 M_\odot \over M} \right) \ \mathrm{s}^{-1} \ ,
\ee
and for the bulk viscosity we find 
\be
{ 1 \over \tau_\mathrm{bv}}  \approx 3.8 \times 10^{-31} \left( { \zeta_\mathrm{eff} \over 1\ \mathrm{g/cms} } \right) \left( { R\over 10\ \mathrm{km} } \right)^5 \left( { 1.4 M_\odot \over M} \right)
\left( { 1\ \mathrm{ms} \over P} \right)^4 \ \mathrm{s}^{-1} \ .
\label{tbv}\ee
Finally, in the case of the mutual friction we find 
\be
{1 \over \tau_\mathrm{mf}} \approx 0.17 \mathcal{B}  \left( { R\over 10\ \mathrm{km} } \right)^4  \left( { P \over 1\ \mathrm{ms}} \right)^{-5} \ \mathrm{s}^{-1} \ .
\label{tmf} \ee
This result differs significantly from the estimate in \cite{mana1}, in particular, in terms of the scaling with the rotation rate. However, our result 
is obtained from the actual counter-moving part of the r-mode solution and the final scaling accords with the standard result (see, for example, \cite{rmode}). 
The  main difference between our result and that in \cite{mana1} can be explained by noting that, if we take the r-mode energy to be  proportional to $w_{\n\s}^2$ rather than $v_\n^2$
then we arrive at a mutual friction damping time scale very similar to that obtained in \cite{mana1}. The large discrepancy between the estimates 
thus follows from the fact that the velocites $v_\n^i$ and $w_{\n\s}^i$ associated with the r-mode solution 
are of different orders of $\Omega$ in the slow-rotation approximation.

\begin{figure}[t]
\centering
\includegraphics[height=8cm,clip]{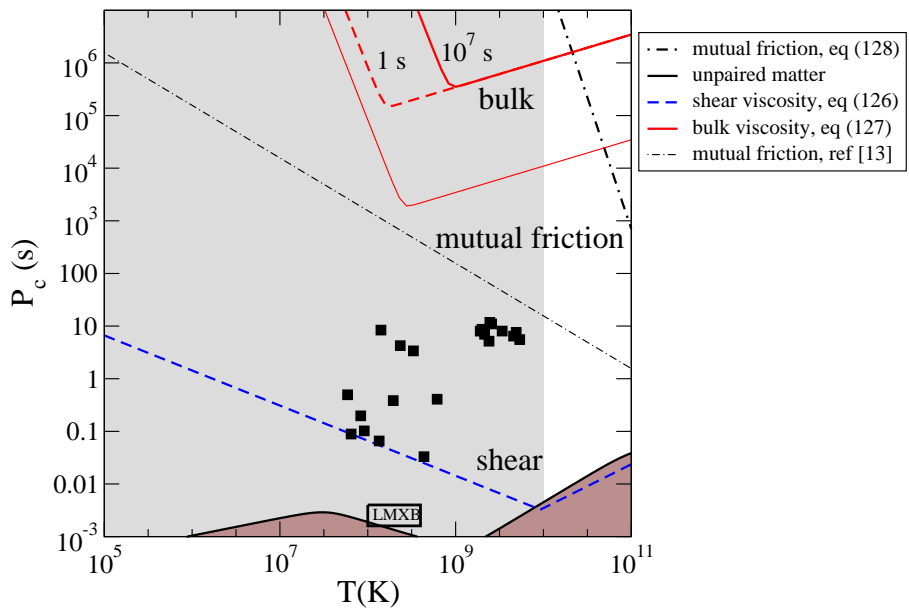}
\caption{The r-mode instability window for a dense core comprising a  CFL superconducting quark condensate and
(finite temperature) phonons. The figure shows the critical rotational period vs core temperature for the instability. 
Gravitational waves drive the $l=m=2$ r-mode unstable for systems located \underline{below} the different curves. The 
effect of shear- and bulk 
viscosity associated with phonon processes are shown as blue (dashed) and red (solid/dashed) curves, respectively.
We show the result for both the canonical value $\beta = 1$ (solid red line) and the rather extreme value $\beta = 10^{-7}$ (dashed red line) in \eqref{ts}, leading to 
relaxation times of $10^7$~s and $1$~s (as indicated in the figure), respectively. 
We also show the result for the case when $\bar{\alpha}$ is increased by a factor of $10^4$ (thin solid red line). 
Our result for the mutual friction damping (this dash-dot black line) is compared to the result from \cite{mana1} (thin dash-dot black line), demonstrating the dramatic effect of the 
different scaling with the rotation.
For comparison, we also provide the instability curve for the combined shear- and bulk viscosity, obtained from \eqref{unpeta} and \eqref{unpzeta}, 
in the case of unpaired quark matter (black line at bottom of figure).  Finally, we  show the observed spin period and inferred core temperature for a number of astrophysical systems (filled squares), 
and we also indicate the region of parameter
space where accreting neutron stars in low-mass X-ray binaries (LMXB) are thought to be located. The region below $10^{10}$~K, where the phonon-fluid model may not be appropriate,  
is shown with grey background to emphasize the need for improved modelling in the astrophysically relevant part of parameter space.}
\label{instab}
\end{figure}

Combining these estimates with \eqref{eta0}, \eqref{etaeff}, \eqref{zeta30} and \eqref{Bcoff}, i.e. balancing  the four timescales \eqref{tgw}--\eqref{tmf},
we arrive at the r-mode instability window in Figure~\ref{instab}.
The figure shows the critical rotational period  for the instability as function of the core temperature for a canonical star with mass $1.4\ M_\odot$ and radius $10$~km. 
Because of the less efficient damping mechanisms, we show the period rather than the rotation rate. This makes sense because the long timescales involved beome much clearer than 
when the critical rotation rate is expressed as a small fraction of the rotational break-up limit. 
As a result, 
gravitational waves drive the $l=m=2$ r-mode unstable for systems located \underline{below} the different curves in the figure. 
Our estimates show that the 
bulk viscosity is much weaker than the shear viscosity at all temperatures of interest. In order to show how the bulk viscosity result depends on
the relaxation time, c.f. \eqref{ts}, we show the result for both the canonical value $\beta = 1$  and the rather extreme value $\beta = 10^{-7}$, leading to 
relaxation times of $10^7$~s and $1$~s, respectively. 
The fact that the associated instability curves do not differ much shows that the results are not very sensitive to this parameter. 
This is due to the strong scaling with temperature. In order to explore the importance of the overall strength of the bulk viscosity, 
we also show the result for the case when $\bar{\alpha}$ is increased by a factor of $10^4$ compared to our canonical value $\bar{\alpha}=1$. This result 
shows that the bulk viscosity has to be vastly different from our assumed model in order to be relevant. The data illustrated in figure~\ref{instab}
also shows that the instability curve obtained from balancing the gravitational-wave driving
and our result for  the phonon mutual friction damping is vastly different from the result in \cite{mana1}. The different dependence on 
rotation is obvious, and it is clear that the phonon mutual friction would be irrelevant for all astrophysical stars with CFL cores. 
For comparison, we  show the instability curve for the combined shear- and bulk viscosity, obtained using \eqref{unpeta} and \eqref{unpzeta}, 
in the case of unpaired quark matter. It is interesting to note that the phonon shear viscosity in the CFL case leads to the strongest r-mode damping at 
temperatures above $10^{10}$~K. Of course, astrophysical neutron stars will cool to temperatures much lower than this soon after birth so this 
region may not be that relevant. The results in the figure show, quite clearly, that we need to improve our understanding of the 
low-temperature regime. The instability curve below the $10^{10}$~K is obtained using our phenomenological model \eqref{etaeff}. 
Future work needs to model this regime in detail.

For comparison, we also show the observed spin and inferred core temperature for a number of astrophysical systems in Figure~\ref{instab} (the data is taken from \cite{pons}). 
This comparison is somewhat inconsistent because we have assumed that the core is shielded by a normal crust with a thickness of, at least, a few hundred meters
(so that the standard heat-blanket argument \cite{gudmund} applies. The presence of a normal matter outer region should provide addition r-mode 
dissipation channels that we have not considered. In particular, one would expect a viscous boundary layer at the crust-core interface
to be a more efficient damping agent (see \cite{ekman1,ekman2} for the most recent discussion) than the phonon mechanisms that we have considered. We also indicate the region of parameter
space where accreting neutron stars in low-mass X-ray binaries would be located (given the usual arguments \cite{bildsten,aks}). These systems would  be 
at variance with the pure CFL plus phonon model in the sense that they would be located deep into the unstable region. 
If there were an active r-mode instability in these systems, one would expect it to have observational effects on, for example, the spin evolution.
Observations do not provide any evidence of this.

\section{Concluding remarks}

We have presented the first true multi-fluid analysis of a dense neutron star core with a deconfined, colour-flavour-locked superconducting, quark core. 
By focussing on a cool system, and accounting only for the condensate and (finite temperature) phonons, 
we  made progress by taking over much of the formalism from the analogous problem for superfluid $^4$He, the archetypal two-fluid 
laboratory system. The additional fluid degree of freedom, in the present case represented by the phonon gas, leads to the system not being well 
represented by the Navier-Stokes equations. In particular, a complete model requires a number of additional viscosity coefficients. 

Without an actual calculation it is not easy to establish whether the multi-fluid aspects are relevant or not. For example, in the case
of the gravitational-wave driven instability of the f-modes it is known that the superfluid degree of freedom is very important, since 
the vortex mutual friction may completely suppress that instability below a critical temperature (see \cite{lm1,fmode}).
It is known that, because of the different nature of the associated velocity field, the mutual friction does not affect the r-mode instability 
in the same drastic fashion \cite{rmode,lm}. These  examples provide clear evidence that different problems need to be considered on a case-by-case basis. 

We have provided a detailed dissipative formulation for a system comprising a quark (CFL) condensate and phonons. The model builds on recent improvements
in our understanding of the analogous problems of superfluid $^4$He \cite{helium} and causal heat conductivity \cite{heat}. A key ingredient 
is the massless entropy component that represents the phonon excitations. We have discussed how the superfluid constraint of irrotationality reduces the 
number of required viscosity coefficients to four (one shear and three bulk), and provided a translation between  results in the literature
\cite{manuel1,manuel2} and the coefficients in our formalism. We also emphasised that many more dissipative channels may come into play 
in a rotating system where superfluid vortices are present \cite{monster,helium}. In particular, we showed how the vortex mediated mutual friction is accounted for in the model, and 
translated the available results for the associated coefficients \cite{mana1}. 

In order to be able to make relevant estimates for the r-mode instability, we developed a simple two-component equation of state based on 
the MIT bag model at zero temperatures  with an additional phonon gas representing the thermal component. This example highlights the additional information that is needed in a
multi-fluid analysis, in particular, regarding the entrainment coupling. Future work needs to provide a consistent equation of state, including \underline{all} the key aspects.  
The model equation of state completed the formulation of the problem and we could, in principle, have carried out a numerical study of the r-modes.
We opted not to do this, instead introducing a sequence of simplifying assumptions, because we felt that it would be useful to start by working out some less precise estimates for the relevant 
dissipation timescales. A numerical analysis of the problem should, of course, be encouraged. The problem has a number of interesting aspects, and 
may shed light on how one should deal with finite temperature superfluid neutron stars in general. 

This work was motivated by a desire to understand the different phases of CFL matter from a hydrodynamics point of view. It is, obviously, an 
interesting problem and the notion that observations of gravitational waves from relativistic stars may help shed light on the extreme QCD phase diagram is exciting. 
The r-mode instability has been discussed in this context for some time, see for example \cite{manuel1,manuel2,mana1,alford1,alford2,madsen,strange}, but there has not been any previous discussion of the 
multi-fluid aspects of the problem. We hope that this work will stimulate a more detailed discussion between experts in the relevant areas. 

The fact that 
the various phonon processes that we have accounted for can be shown to have little effect on the r-mode instability should not discourage 
future efforts. In fact, the result could probably have been anticipated. The real challenge will be to account for additional degrees of freedom, 
especially associated with the kaons (either thermal or in a condensate) \cite{alford1,alford2}, and the modelling of ``hybrid'' stars with quark cores and various phase transitions. 
These problems have additional features that, while possibly
understood in principle, have never been considered in practice. This makes the modelling more complex, but also  intriguing since
the richer dynamics may  lead to surprises.  

\section*{Appendix: Translation to the orthodox framework}

The relationship between the flux conservative formulation and the orthodox framework for superfluid Helium has  been examined  in detail elsewhere \cite{helium,prix}. 
Still, it is useful to summarise the main results that are needed to relate the different dissipation coefficients.
First of all, we identify  the ``normal'' fluid in the standard description with the gas of excitations, which is directly associated with the entropy of the system. This leads to
\be
v^i_\N=v^i_\s \ .
\ee
Secondly, the so-called ``superfluid'' velocity is given by
\be
v^i_\S=\frac{\pi^i_{\n}}{\rho}=(1-\varepsilon)v_\n^i+\varepsilon v_\s^i \ ,
\ee
where $\varepsilon=\frac{2\alpha}{\rho}$ is the entropy entrainment parameter.
This means that the total mass flux  takes the form
\be
\rho v_\n^i=\rho_\N v_N^i+\rho_\S v^i_\S \ ,
\ee
from which we learn that
\be
\rho_\S = \frac{\rho}{1-\varepsilon} \quad \mbox{and} \quad  \rho_\N=-\frac{\varepsilon \rho}{1-\varepsilon} \ ,
\ee
where $\rho_\S$ and $\rho_\N$ are the superfluid and normal fluid densities. 

\acknowledgements
We are grateful to Mark Alford and Cristina Manuel for a number of useful discussions.
NA and BH acknowledge support from STFC via grant number PP/E001025/1.
BH also acknowledges support from the European Science Foundation (ESF) for the activity entitled ``The New Physics of Compact Stars'' , under exchange grant 2449, 
and thanks the Dipartimento di Fisica, Universit\`a degli studi di Milano for kind hospitality during part of this work.


\begin{references}

\bibitem{khalatnikov} I.M. Khalatnikov, {\em An introduction to the theory of
superfluidity} (W. A. Benjamin, Inc., New York, 1965).

\bibitem{putterman} S.J. Putterman, {\em Superfluid hydrodynamics}
(North-Holland, Amsterdam, 1974).

\bibitem{epstein} R.I. Epstein, Ap. J. {\bf 333},  880 (1988).

\bibitem{mendell} L. Lindblom \& G. Mendell, Ap. J. {\bf 421} 689 (1994).

\bibitem{ac01} N. Andersson \& G.L. Comer, MNRAS {\bf 328} 1129 (2001).

\bibitem{gusakov} M.E. Gusakov \& N. Andersson, MNRAS {\bf 372}, 1776 (2006).

\bibitem{NPA} N. Andersson, G.L. Comer \& K. Glampedakis, Nucl. Phys. A {\bf 763}, 212 (2005).

\bibitem{CFL} M. Alford, A. Schmitt, K. Rajagopal \& T. Sch\"afer, Rev. Mod. Phys. {\bf 80}, 1455 (2008).

\bibitem{helium} N. Andersson \& G. L. Comer,  Entropy entrainment in finite temperature superfluids, preprint arXiv:0811.1660

\bibitem{jaikumar} P. Jaikumar, G. Rupak \& A.W. Steiner, Phys. Rev. D {\bf 78} 123007 (2008).

\bibitem{manuel1}  C. Manuel \& F.J. Llanes-Estrada, JCAP {\bf 8}, 1 (2007).

\bibitem{manuel2} C. Manuel, A. Dobado \& F.J. Llanes-Estrada, JHEP { \bf 9}, 76 (2005).

\bibitem{mana1} M. Mannarelli, C. Manuel \& B.A. Sa'd, Phys. Rev. Lett. {\bf 101},  241101 (2008).

\bibitem{mana2} M. Mannarelli \& C. Manuel, Phys. Rev. D {\bf 81}, 043002 (2010).

\bibitem{alford1} M.G. Alford, M. Braby, S. Reddy \& T Sch\"afer, Phys. Rev. C. {\bf 75} 055209, (2007).

\bibitem{alford2} M.G. Alford, M. Braby \& A. Schmitt,  J. Phys. G.  {\bf  35}, 115007 (2008).

\bibitem{alford3} M.G. Alford, M. Braby \& S. Mahmoodifar, Phys. Rev. C {\bf 81} 025202 (2010).

\bibitem{prix} R. Prix, Phys. Rev. D {\bf 69} 043001 (2004).

\bibitem{monster} N. Andersson \& G.L. Comer, Class. Quantum Grav. {\bf 23}
5505 (2006).

\bibitem{livrev} N. Andersson \& G.L. Comer, Living Reviews in Relativity, {\bf 10} no. 1 (2007).

\bibitem{heat}  N. Andersson \& G.L. Comer, Proc. R. Soc. London A, doi 10.1098/rspa.2009.0423

\bibitem{trev} N. Andersson, T. Sidery \& G.L. Comer, MNRAS {\bf 368}, 162 (2006).

\bibitem{alcock} C. Alcock. E. Farhi \& A. Olinto, Ap. J. {\bf 310}, 261 (1986).

\bibitem{review} N. Andersson \& K.D. Kokkotas, Int. J. Mod. Phys. D {\bf 10},  381 (2001).

\bibitem{hyperon} B. Haskell \& N. Andersson, Superfluid hyperon bulk viscosity and the r-mode instability of rotating neutron stars, preprint arXiv:1003.584

\bibitem{kl1} K.H. Lockitch, N. Andersson \& J.L. Friedman, Phys. Rev. D {\bf 63},  024019 (2001).

\bibitem{kl2} K.H. Lockitch, J.L. Friedman  \& N. Andersson, Phys. Rev. D {\bf 68},  124010 (2003).

\bibitem{jr1} J. Ruoff \& K.D. Kokkotas, MNRAS {\bf 328},  678 (2001).

\bibitem{jr2} J. Ruoff \& K.D. Kokkotas, MNRAS {\bf 330}, 1027 (2002).

\bibitem{greg} G.L. Comer, Found. Phys. {\bf 32} 1903 (2002). 

\bibitem{fmode} N. Andersson, K. Glampedakis \& B. Haskell, Phys. Rev. D {\bf 79}, 103009 (2009).

\bibitem{rmode} B. Haskell, N. Andersson \& A. Passamonti, MNRAS {\bf 397}, 1464 (2009).

\bibitem{prix2} R. Prix \& M. Rieutord, Astron. Astrophys. {\bf 393}, 949 (2002).

\bibitem{passamonti} A. Passamonti, B. Haskell \& N. Andersson, MNRAS {\bf 396}, 951 (2009).

\bibitem{cl95} B. Carter \& D. Langlois, Phys. Rev. D {\bf 51} 5855 (1995).

\bibitem{helsound} A.A.	Zadorozhko, E. Ya. Rudavski, V.K. Chagovets, G.A. Sheshin \& Yu. A. Kitsenko, Low. Temp. Phys.  {\bf 35}, 100 (2009).

\bibitem{madsen} J. Madsen, Phys. Rev. D {\bf 46}, 3290 (1992).

\bibitem{book} R.J. Donnelly, {\em Quantized vortices in Helium II} (Cambridge University Press, Cambridge, 1991).

\bibitem{prec} K. Glampedakis, N. Andersson \& D.I. Jones, MNRAS {\bf 394}, 1908 (2009).

\bibitem{lom} L. Lindblom, B.J. Owen \& S.M. Morsink, Phys. Rev. Lett. {\bf 80}, 4843 (1998)

\bibitem{lmo} L. Lindblom \& G. Mendell, Phys. Rev. D {\bf 60}, 064006 (1999).

\bibitem{pons} D.N. Aguilera, J.A. Pons, J.A. Miralles, Ap. J. Lett. {\bf 673}, 167 (2008).

\bibitem{gudmund} E.H. Gudmundsson, C.J. Pethick \& R.I. Epstein, Ap. J.  {\bf 272},  286 (1983).

\bibitem{ekman1} K. Glampedakis \& N. Andersson, MNRAS {\bf 371},  1311 (2006).

\bibitem{ekman2} K. Glampedakis \& N. Andersson, Phys. Rev. D {\bf 74}, 044040 (2006).

\bibitem{bildsten} L. Bildsten, Ap. J. Lett. {\bf 501}, 89 (1998). 

\bibitem{aks} N. Andersson, K.D. Kokkotas \& N. Stergioulas, Ap. J. {\bf  516}, 307 (1999).

\bibitem{lm1} L. Lindblom \& G. Mendell, Ap. J.  {\bf 444}, 804 (1995).

\bibitem{lm} L. Lindblom \& G. Mendell, Phys. Rev. D {\bf  61}, 104003 (2000).

\bibitem{strange} N. Andersson, D.I. Jones \& K.D. Kokkotas, MNRAS {\bf 337},  1224 (2002).

\end{references}
\end{document}